\documentclass[english]{emulateapj}
\makeatletter

\providecommand{\tabularnewline}{\\}

\shorttitle{Constraints on the origin and nature of HVSs}
\shortauthors{Perets, H. B.}

\makeatother

\begin{document}
\newcommand{\Mo}{M_{\odot}}
\newcommand{\Ro}{R_{\odot}}
\newcommand{\Lo}{L_{\odot}}
\newcommand{\SgrA}{\mathrm{Sgr\, A^{\star}}}
\newcommand{\Ms}{M_{\star}}
\newcommand{\Mbh}{M_{\bullet}}
\newcommand{\rMP}{r_{\mathrm{MP}}}
\newcommand{\aGW}{a_{\mathrm{GW}}}

\title{Dynamical and evolutionary constraints \\
on the nature and origin of hypervelocity stars }

\author{Hagai B. Perets }

\email{hagai.perets@weizmann.ac.il}
\affil{Weizmann Institute of Science, POB 26, Rehovot 76100, Israel}

\begin{abstract}
In recent years several hypervelocity stars (HVSs) have been observed
in the halo of our Galaxy. Such stars are thought to be ejected through
dynamical interactions near the massive black hole (MBH) in the Galactic
center. Three scenarios have been suggested for their ejection; binary
disruption by a MBH, scattering by inspiraling IMBH and scattering
by stellar BHs close to MBH. These scenarios involve different stellar
populations in the Galactic center. Here we use observations of the
Galactic center stellar population together with 
dynamical and evolutionary arguments
to obtain strong constraints on the nature and origin of HVSs. 
We show that the IMBH
inspiral scenario requires too many ($\mathcal{O}(10^{3}$) main sequence
B stars to exist close to the MBH ($<0.01$ pc) at the time of inspiral,
where current observations show $\mathcal{O}(10)$ such stars. Scattering
by SBHs also require too many B stars to be observed in the GC, but
it may contribute a small fraction of the currently observed HVSs.
The binary disruption scenario is still consistent with current observations.
In addition it is shown that
recently suggested signatures for HVSs origin such as hypervelocity
binaries and slow rotating HVSs are much weaker than suggested
and require too large statistics. 
In addition, we show that due to the conditions close to the MBH most
binary star systems are not expected to survive for long in this region.
Consequently, unique stellar populations that require long evolution
in binaries are not expected to be ejected as HVSs in the BHs scattering
mechanisms (this may also be related to to the recently observed asymmetry
in the velocity distribution of HVSs). 
\end{abstract}


\section{Introduction}

In recent years several hypervelocity stars (HVSs) have been observed
in the halo of our Galaxy, some of them unbound to the Galaxy \citep{bro+07b}.
Most of the observed HVSs are B-type stars \citep{bro+05,bro+06a,bro+06b,bro+07a,bro+07b,ede+06},
implying a Galactic population of $96\pm10$ such unbound HVSs (closer
than $\sim120\, kpc$ to the Galactic center (GC); \citealt{bro+07b}).
Given the color selection of the targeted survey for these stars \citep{bro+06a},
such stars could be either main sequence (MS; or blue straggler) B
stars or hot blue horizontal branch (BHB) stars. Only four of the observed
B type stars have stellar type identification, and were found
to be B type MS stars \citep{ede+06,fue+06,prz+08,lop+08}. In addition a single
subdwarf O HVS has been observed \citep{hir+05}. More recently a new
HVSs survey of A type stars have detected additional HVSs \citep{bro+08}. We also note 
that more HVSs may be detected in the future in M31 \citep{she+07} and other
galaxies.

Extreme velocities as found for these stars most likely suggest a
dynamical origin from an interaction with the massive black hole (MBH)
in the GC. Several scenarios have been suggested for ejection of HVSs;
a disruption of a stellar binary by the MBH in the GC (\citealt{hil88,yuq+03,per+07};
hereafter the binary disruption scenario), an interaction of a single
star with an intermediate mass black hole (IMBH) which inspirals to
the GC (\citealt{han+03a,yuq+03,lev05}; hereafter the IMBH inspiral
scenario), or interaction with stellar black holes (SBHs) in the GC
(\citealt{yuq+03,mir+00,ole+07}; hereafter the SBHs kicks scenario).
The later two scenarios scatter HVSs mostly from regions very close
to the MBH ($<0.01$ pc) where as the binary disruption scenario mostly
eject HVSs that evolved in binaries much further from the MBH ($\gtrsim2$
pc). The IMBH inspiral scenario is a discrete event, which does not
occur continuously (although a sequence of several IMBH inspirals
may eject HVSs semi-continuously; \citealt{loc+07}) where the binary
disruption or SBHs kicks are continuous processes leading to a constant
rate of HVSs ejection. The different stellar populations involved
in the different scenarios, the importance of binarity in the binary
disruption scenario and the dynamical history of HVSs ejection could
thus help to constrain the nature and origin of HVSs. 

Recently several methods were suggested for discriminating between
the HVSs ejection mechanisms. These include the differences in the
velocity and directional distribution of HVSs \citep{lev05,bau+06,ses+07a},
the binarity of HVSs \citep{luy+07} and the rotational velocities
of HVSs \citep{han07}. \citet{bro+07b}, \citet{sve+07} and \citet{ken+08} 
discussed
the propagation of observed HVSs and the asymmetric distribution of
ingoing and outgoing HVSs (with negative and positive radial velocities,
respectively, in Galactocentric coordinates) in regard to their nature
(MS B stars or hot BHB stars). 

Here we use the current observations of HVSs, observations of the
stellar population in the GC, and dynamical arguments to further constrain
the possible scenarios for the origin of HVSs. We show that the population
of B type MS stars required by the IMBH inspiral scenario and the
SBHs kicks scenario are too large to be consistent with current observations.
We then discuss some unique stellar populations that require long
evolution in binaries, and suggest that they are not likely to be
ejected as HVSs in the SBHs kicks or IMBH inspiral scenarios, since
their binary progenitors are not likely to survive in the harsh environment
close to the MBH. We also discuss the implications of the short survival
time of binaries to the distribution of HVSs rotational velocities and
show that these are not likely to serve as
good discriminators for HVSs origin.
Finally we shortly discuss how these arguments may be related to the
recently observed asymmetric velocity distribution of observed HVSs
\citep{bro+07b}.

\section{Constraints from the young stellar population in the Galactic center}

In each of the HVSs ejection scenarios the unbound HVSs reflect only
a fraction of the total number of stars ejected from the GC. Many
more stars are ejected at lower velocities, but high enough to escape
the close environment of the MBH. Given the inferred number of B type
HVSs from current observations one can obtain the total number of
such ejected stars. Therefore we can infer the number of such stars
that have existed in the appropriate environment of the GC during
the time period of HVSs ejection. In the following we consider the
constraints on the HVSs scenarios suggested by such considerations.
We assume a total number of unbound HVSs of $\sim100$ \citep{bro+07b}.
This is probably only a lower limit for the total number of HVSs,
since many of them might have had higher velocities and propagated
beyond the $\sim120\, kpc$ currently observed, therefore the constraints
suggested here might be more stringent .

\subsection{The IMBH inspiral scenario }

\label{sec:IMBH-scenario}

In the IMBH inspiral scenario \citep{han+03a,yuq+03}, an IMBH inspirals
to the Galactic center through dynamical interactions with stars.
In the late stages of the inspiral, when the IMBH is already close
to the MBH in the GC ($<0.01$ pc or even less, depending on the IMBH
mass), it may closely interact with stars and scatter them at very
high velocities, thus producing HVSs \citep{lev05,bau+06,loc+07,ses+07b}.
Consequently the population of HVSs ejected by an IMBH in the GC should
be strongly correlated with the stellar population in the central
$0.01$ pc of the GC%
\footnote{One could suggest that the ejected stars do not belong to the cusp
population near the MBH, but have inspiralled in the cluster accompanying
the IMBH, and have been scattered during the inspiral. However, simulations
of such scenario show that the young stars are stripped from the cluster
much before the IMBH reaches the central 0.01 pc (see e.g. \citealt{lev+05,ber+06b},
see also observational evidence against the IMBH inspiral origin for
the young stars in the GC \citealt{pau+06}) from which region HVSs
can be ejected in this scenario. %
}. In this scenario the stellar type of the stars (and hence their
mass) only negligibly affect the possibility of their ejection as
HVSs. Therefore any star in the close region of the MBH may become
a HVS in this case. 

Results of N-body simulations \citep{bau+06,loc+07,ses+07b} show
that only a fraction of the stars are ejected as HVSs, and most are
ejected at lower velocities. \citet{ses+07b} find a total number
of 500 (2500) HVSs ejected for an IMBH mass of $\sim5\times10^{3}\, M_{\odot}$($\sim1.5\times10^{4}\, M_{\odot}$)
. They find the total number of stars ejected during the inspiral
to be about $10^{4}$ ($3.5\times10^{4}$). These numbers indicate
that a fraction of $f_{HVS}\sim0.05$ ($0.07$) of all ejected stars
during the IMBH inspiral are HVSs. Current observations infer $\sim100$
unbound B-type HVSs exist in the galaxy, and therefore $\sim100/f_{HVS}\sim2000$
($1400$) B-type stars must have been existed at distance of $0.001$
($0.01$) pc from the MBH during the short time of IMBH inspiral.
Current observations show $\sim1$ ($30$) B-type stars at $0.001$
($0.01$) pc from the MBH (\citep{eis+05,ghe+05}; less luminous B-type
stars might be missed so this could get a factor of $3-8$ times larger%
\footnote{Assuming a present day mass function, and assuming the B stars to
have masses between $3-15\, M_{\odot}$ the relative fractions of
the observed (more luminous) B stars in the overall population of
B stars can be estimated. The fraction of such stars $f_{4-15}$or
$f_{5-15}$ with masses between $4\, M_{\odot}<m<15\, m_{\odot}$
or $5\, M_{\odot}<m<15\, m_{\odot}$, respectively, are $f_{4-15}=0.31$
and $f_{5-15}=0.13$, i.e. up to $\sim3-8$ times more B stars than
observed (with lower masses) could be missed in these regions. In
these calculations we assumed star formation at a constant rate with
a Miller-Scalo IMF \citep{mil+79}, and use a stellar population synthesis
code \citep{ste+03} with the Geneva stellar evolution tracks \citep{sch+92a}
to estimate that the present day number fraction of stars in the S-star
mass range. Using an initial mass function instead gives a smaller
factor of only $2-3$ times undetected B stars.%
}. One may suggest that the stellar population in the GC at the time
of inspiral (up to $\sim10^{8}$yrs ago) was significantly different
than currently observed, however the possibility of so many B-type
stars existing in this small region of the GC is highly unlikely,
and would require a yet unknown process for producing such overabundance
of B-type stars. Given our theoretical understanding of this process,
the current observations of the GC and the inferred number of HVSs,
the IMBH inspiral scenario is unlikely to be the main origin for the
ejection of currently observed HVSs.

\subsection{The SBHs kicks scenario}

In the SBHs kicks scenario \citep{mir+00,ole+07}, SBHs in the close
environment of the MBH (mostly $<0.01$ pc or even $<0.001$ pc; \citet{ole+07})
strongly interact with stars and scatter them at high velocities thus
producing HVSs. Consequently, and similar to the IMBH scenario, the
population of HVSs ejected by an IMBH in the GC should be strongly
correlated with the stellar population in the central $0.01$ pc of
the GC. In this scenario the stellar type of the stars (and hence
their mass) affects the possibility of their ejection as HVSs, but
not strongly for the B MS stars of $3-4$$M_{\odot}$ currently observed. 

Results of analytic calculations by \citet{ole+07} indicate that
only a fraction of the stars are ejected as HVSs, and most are ejected
at lower velocities. They find that the ejection rate of stars at
lower velocities than required for HVSs ($\lesssim800km\, s^{-1}$)
goes like $(v_{ej}/800\, km\, s^{-1})^{-2.5}$. For the $100$ unbound
HVSs inferred from observations about $\sim100\times(100/800)^{-2.5}\simeq1.8\times10^{4}$
B type stars have been ejected from the central $0.01$ pc with $>100\, km\, s^{-1}$,
i. e. for the lifetime of such $3-4\, M_{\odot}$ stars, a star formation
rate of at least $\sim2\times10^{4}/2\times10^{8}=10^{-4}$ B stars
per year is required. Given that most of the stars are not scattered
outside the GC region, this is only a lower limit on the required
star formation rate, and at least as many stars should have been left
over in this region. 

It is unlikely that regular star formation could have formed such
stars so close to the MBH, given the tidal forces in this region (that
would disrupt a progenitor molecular cloud). These stars might have
formed continuously through a fragmentation of a gaseous disk, although
so close to the MBH even such star formation would most likely be
prohibited or be inefficient given the required Toomre criteria in
this region \citep{lev07}. Nevertheless, these stars could have formed
at some larger distance such as the young stars observed at $<0.5$
pc scale stellar disk in the GC \citep{pau+06} and continuously migrated
close to the MBH \citep{lev07}. Such scenario would require the star
formation rate at this region (most likely $<0.5$ pc) to be $>10^{-4}\, yr^{-1}$,
and would also require an efficient mechanism for transferring a large
fraction of these formed stars to the central $0.01$ pc at short
times. The relaxation time in the GC is much longer than the lifetime
of such stars and therefore some other migration mechanism would be
required (e.g. migration in a gaseous disk; \citealt{lev07}). The
central region of the GC probably contains $\lesssim5\times10^{2}$
B type stars such similar to the observed HVSs (Eisenhauer, F.; private
communication (2006)), and therefore implies a star formation rate
of $<5\times10^{2}/2\times10^{8}=2.5\times10^{-6}$ such B stars per
year. This rate is much smaller than the minimally required rate of
$\sim10^{-4}\, yr^{-1}$ we found above for explaining the inferred
number of HVSs in this scenario. More simply put we would have expected
to observe $\sim10^{4}$ B stars in the GC region where only $\sim500$
are observed  (or even less in the central 0.01 pc), suggesting that 
this scenario is unlikely to be the
main origin for the ejection of currently observed HVSs, although
it could explain a fraction of them.

\subsection{The binary disruption scenario}

In the binary disruption scenario \citep{hil88} binaries are disrupted
by the MBH in the GC if they come closer than the tidal radius. One
star is captured by the MBH where the other is ejected at high velocity
thus producing HVSs. The fraction of ejected stars with velocities
lower than those of HVSs is strongly dependent on the the semi-major
axis distribution of the binaries (where higher velocity stars are
ejected from disruption of closer binaries \citep{hil91,bro+06c})
which is unknown in the GC. The velocity of an ejected star was found
in numerical simulations \citet{hil88,bro+06c} to scale as\begin{eqnarray}
v_{\mathrm{BH}} & = & 892\,\mathrm{km\, s^{-1}}\times\nonumber \\
 &  & \left(\frac{a}{1\,\mathrm{AU}}\right)^{-1/2}\left(\frac{M_{bin}}{8M_{\odot}}\right)^{1/3}\!\left(\frac{M_{BH}}{3.7\!\times\!10^{6}\,\Mo}\right)^{1/6}\,.\end{eqnarray}
 To reproduce the high HVS velocities we consider binaries with $a\!<\!0.95$
AU. These are tidally disrupted at $r_{t}\!<\!3.8\times10^{-4}$ pc
and eject an unbound HVS with $v_{\mathrm{BH}}\!\gtrsim\!920\,\mathrm{km\, s^{-1}}$
which could be observed as an HVS with velocity $>450\, km\, s^{-1}$
at 55 kpc from the GC, given estimated Galactic potential difference
between the center and 55 kpc of $v_{55}\!\sim\!800\,\mathrm{km}\,\mathrm{s}^{-1}$
\citep{car+87}). 

For the semi-major axis distribution of massive binary stars, which
is strongly biased towards close binaries a large fraction of all
binaries, $(f_{bin}\sim0.3-0.9)$ have semi-major axis short enough
($\lesssim AU$; \citealt{gar+80,abt83,kob+06}) , such that the binary
disruption by the MBH would lead to a ejection of HVS. Therefore given
the $\sim100$ HVSs inferred from observations one would require $\sim330/(f_{bin}/0.3)$
binaries to be disrupted. This does not constrain the stellar population
from which the binaries originate (most originate from the central
$10\, pc$ of the GC where $\sim10^{4}$such binaries exist), but
may constrain the number of captured stars \citep{per+07}. In each
binary disruption the companions to the ejected stars are captured
by the MBH. The capture semi-major axis distance to the MBH is linearly
dependent on the semi-major axis of the original stellar binary \citep{hil91},
which is $\lesssim0.02$ pc for the companion of a HVS) and therefore
$100-300$ such stars should be captured near the MBH during the last
$\sim10^{8}$yrs in this region. This is generally consistent with
current observations of $\sim100$ massive B stars at $<0.04$ pc
from the MBH, where the number of less massive (less luminous) B stars
may be a few times larger$^{2}$. 

Although less likely, the initial semi-major axis distribution of
B MS binaries in the GC environment may behave like that of lower
mass stars \citep{duq+91}. Most binaries with large semi-major axis
would not survive for long in the central regions of the GC ($<10$
pc; from which most disrupted binaries originate; see fig. \ref{f:evap_time}),
and so only the closer binaries (about half of the primordial population)
survives. In this case the fraction of disrupted binaries which lead
to ejection of HVSs (binaries closer than $\lesssim AU$) is $\sim0.16$.
Therefore we obtain a total number of $\sim100/0.16=625$ stars captured
by the MBH during the last $\sim10^{8}$yrs. However, many of these
captured stars would be captured at much larger distances than the
companions of HVSs and would be distributed up to distances of $\sim pc$
from the MBH; i.e. the constrain we have is of $\sim625$ B type stars
(more likely a few times more, if both stars are captured, and also
taking into account the larger impact parameter for wider binaries)
in the central pc. This is still marginally consistent with current
observations in this region, but might be excluded with future observations.
We conclude that the scenario of HVS ejection from binary disruption
is consistent with current observations of B stars in the GC given
a binary distribution of B stars binaries similar to that observed
in the solar neighborhood, and marginally consistent if the binary
distribution is similar to that of lower mass stars in the solar neighborhood.

\section{Constraints from the evolution of binary systems in the Galactic
center}

Stellar evolution in binary systems can be very different than the
evolution of isolated stars. In such systems the binary components
may interact in many ways, whether through mass transfer, tidal forces,
winds, radiation or other ways. Such interaction can considerably
change the evolution of the stars and lead to unique characteristics
of stars that are different or not even accessible to stars evolved
in isolation. Some of these effects require long term evolution in
binaries. Other effects are related to the formation process of a
binary system (e.g. stars in binary systems show lower average rotational
velocities than single stars, irrespective of their age \citep{abt+04}).
Observationally, several peculiar stellar populations are observed
mostly or only in binaries \citep{abt83}.

The different evolution of stars in binaries can be used for discriminating
between ejection scenarios of HVSs and help to understand and predict
their characteristics. Recently, two such discriminators have been
suggested. The binary disruption scenario, by definition, involves
the ejection of a single star which evolved in a binary. It was pointed
out that binary components have lower average rotational velocities
\citep{abt+04}, and therefore HVSs from such a scenario should similarly
be slow rotators \citep{han07}. In the inspiraling IMBH scenario
both single and binary stars could be ejected as HVSs. The later possibility
of a binary HVS has been suggested as a unique signature of the IMBH
inspiral scenario \citep{luy+07}. In the following we generalize
the use of binary evolution as a signature of HVSs ejection scenarios
(and predictors for their nature) and suggest additional signatures.
However, we also show that the dynamics of binaries in the GC usually
make this type of signatures only weak signatures at most, and would
probably require large statistics to be useful discriminators in most
cases. Nevertheless, these may better constrain the characteristics
of HVSs ejected from the GC and may help explain the asymmetric velocity
distribution of observed HVSs.

We note that all of the arguments given below are predictors not only
for the characteristics of HVSs, but also for stars observed close
to the MBH, that were either formed close by (e.g. in the recently
observed stellar disk; \citet{pau+06}) or captured through the binary
disruption mechanism \citep{per+07}.

\subsection{Binary survival in the Galactic center}

\begin{figure}
\begin{tabular}{c}
\includegraphics[clip,width=0.9\columnwidth]{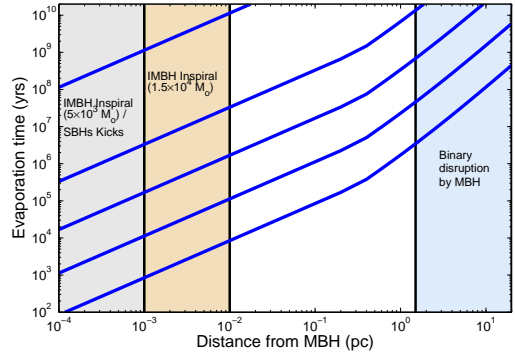}\tabularnewline
\end{tabular}

\caption{\label{f:evap_time}Evaporation time of binaries in the Galactic center
for binaries with different semi major axis; 0.01 AU (contact binaries),
0.1 AU, 1 AU, 10 AU and 100 AU (from top to bottom). The shaded regions
show the distance range from which most hypervelocity stars are ejected
in the IMBH inspiral scenario (IMBH masses of $1.5\times10^{4}M_{\odot}$and
$5\times10^{3}M_{\odot}$;\citet{ses+07b}); the SBHs kick scenario
\citep{ole+07}; and the binary disruption scenario \citep{per+07}. }
 
\end{figure}

Binaries may survive for a Hubble time unless destroyed due to stellar
evolutionary processes (e.g. merger or disruption due to mass transfer
or mass loss) or subjected to dynamical interactions. In dense environments
the later possibility may play an important role in the evolution
of binary systems. In such environments binaries (soft binaries; \citealt{heg75})
may gradually evaporate due to perturbations from encounters with
other single stars if \begin{equation}
|E|/m_{bin}\sigma^{2}<1,\label{eq:hard-soft}\end{equation}
 where $E=-Gm_{1}m_{2}/2a$ is the orbital energy of a binary with
component masses $m_{1}$and $m_{2}$and separation $a$, $m_{bin}=m_{1}+m_{2}$is
the binary mass and $\sigma$ is the velocity dispersion of stars
in the system. Due to the high velocity dispersions in the GC, all
but the closest (contact) binaries are soft binaries. Hard close binaries
can become harder due to interactions with other stars, i.e. become
even closer. However the hardening changes the orbital energy of these
binaries at a rate of less than \textasciitilde{}20 percents per relaxation
time for marginally hard binaries, and even less for harder binaries
(see e.g. Eq. 8-113 in \citealt{bin+87}). This would only slightly
change the distribution of periods of close binaries. Given the uncertainties
in the distribution of binaries in the GC this effect is negligible
and do not contribute much to the processes of binary evolution in
the GC. Most of the binary population in the GC is in soft binaries.
The evaporation time of such binaries is given by (\citep{bin+87}\begin{equation}
t_{evap}=\frac{m_{12}}{m}\frac{\sigma}{16\sqrt{\pi}\rho a\ln\Lambda},\label{eq:tevap}\end{equation}
where $\rho$ is the stellar density, $m$ is the typical mass of
a star in this region and $\ln\Lambda$ is the Coulomb logarithm.
In the GC $\sigma$ is dependent on $r$; $\sigma\sim\sqrt{GM(<r)/r}$,
where $M(>r)$ is the enclosed mass up to distance $r$ from the MBH.
Fig. \ref{f:evap_time} shows the evaporation time for binaries with
different semi-major axis ($10^{-2}-10^{2}\, AU$) in the central
regions in the GC, taking $\rho(r)=\rho_{0}(r/r_{0})^{-\alpha},$
where $r_{0}=0.4$ pc, $\rho_{0}=1.2\times10^{6}M_{\odot}{\rm pc}^{-3}$,
$\alpha=1.4$ for $r<r_{0}$ and $\alpha=2$ for $r>r_{0}$\citep{gen+03a}.
The binary mass ratio is assumed to be 1 $(m_{12}=2m)$. 

%

Typical low mass ($<3M_{\odot}$) binaries have a log normal distribution
of semi-major axis centered around $\sim30\, AU$ \citep{duq+91}.
As can be clearly seen in fig. \ref{f:evap_time} most such binaries
can not survive for long close to the MBH. Many of the peculiar properties
of stars evolved in binaries are due to their long term evolution
in such systems \citep{abt83}. Since binaries close to the MBH are
disrupted in very short time scales, the component stars in these
binaries would become single stars, and effectively evolve as isolated
stars. Consequently, peculiar stellar populations that require long
term evolution in binaries are not expected to form in these regions.
As discussed earlier, the scenarios of HVSs ejection by SBHs or by
an IMBH are most efficient at close distances of $\sim0.001-0.01$
pc from the MBH, and therefore most HVSs are ejected from these region
in these scenarios \citep{ole+07,ses+07b}. At such distances from
the MBH the velocity dispersion is of few$\times10^{2}-10^{3}\, km\, s^{-1}$,
and even the closest binaries are soft and would be disrupted in less
than $10^{7}$yrs (see fig. \ref{f:evap_time}), i.e. shorter than
the main sequence lifetimes of most stars. Consequently, hypervelocity
binaries that were suggested as a possible signature of the IMBH inspiral
scenario \citep{luy+07} and peculiar stellar populations evolved
and observed mainly in binaries are not expected to be ejected as
HVSs in these scenarios (see also \citet{loc+07}). Other stellar
populations include for example subdwarf B (sdB) stars (\citet{max+01,han+03b},
Am stars (see also \citet{han07}) and BY Dra stars (see \citet{abt83}
for a review). 

In the binary disruption scenario for ejection of HVSs a different
picture arises. In this case most binaries disrupted by the MBH come
from much larger distances from the MBH ($\gtrsim2\, pc$; \citet{per+07})
than HVSs ejected in the SBHs kick or IMBH inspiral scenarios. At
these distances binaries could survive longer (fig. \ref{f:evap_time}).
However, a HVS is ejected following the disruption of the binary,
destroying the possible progenitor of any binary evolved peculiar
star. Consequently, only stars that already evolved in a binary to
become peculiar prior to the disruption of the binary could be ejected
as peculiar type HVSs. Unfortunately the lifetime of many peculiar
stars at this phase are usually much shorter than their lifetime on
the MS (e.g. the lifetime of sdB stars are of the order of a $1-2\times10^{8}$yrs
\citep{dor+93}, where their progenitor MS lifetimes could be a few
Gyrs) and therefore fine tuning would be required for the ejection
of HVSs in this case (i.e. they need to be ejected in the short time
after they become peculiar, and before they end their life at this
phase) and they would be rare. Nevertheless, if observed, they are
expected to be single stars, which would be a strong signature of
their binary disruption origin, since such stellar populations are
expected to be and usually observed as binary stars. 

Some stellar populations do not exist in binaries, or exist only in
long period binaries. In the binary disruption scenario, such stellar
populations are not (or rarely) expected to be ejected as HVSs. For
example, Be type stars and A4-F2 type stars are usually observed with
large semi major axis, and their binary fraction at smaller semi-major
axis ($<\sim AU$) is low \citep{abt83,abt+84}. Statistics of this
type of stars in HVSs observations (or of stars very close to the
MBH in the GC, where they could have been captured in the binary disruption
mechanism; see e.g. \citealt{gou+03,per+07}), could give a measure
of this possible signature for the HVSs origin from binary disruptions.

\subsection{Rotational velocities of hypervelocity stars}

Recently it was suggested that the rotational velocity of HVSs can
serve as a signature for their origin \citep{han07}. Observations
show that field A and B type MS stars that evolve in binaries have
lower average rotational velocities than isolated stars \citep{abt+02,abt+04}.
If HVSs origin is from the binary disruption scenario, they are expected
to form in binaries and therefore be slower rotators on average. Lower
rotational velocities have been observed even for relatively young
MS stars in binaries, suggesting that the low rotations are related
to their formation in a binary and are not a consequence of their
later evolution in that system \citep{abt+04}. Consequently stars
formed in binaries should show this signature even if their binaries
have been disrupted in a short time. 

\begin{figure}
\begin{tabular}{c}
\includegraphics[clip,width=0.9\columnwidth]{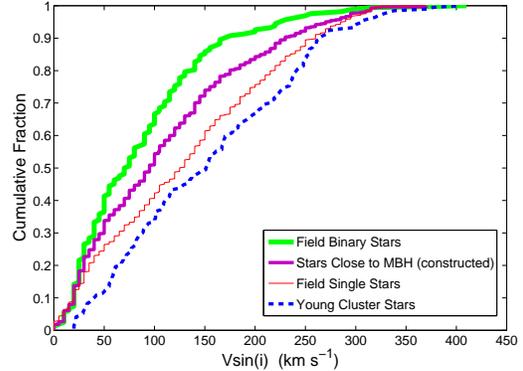}\tabularnewline
\end{tabular}

\caption{\label{f:cdfs}Cumulative distribution of the rotational velocities
of massive B stars of different environments or populations. From
top to bottom, field stars in binaries (\citet{abt+04}; thick solid
line), constructed distribution close to the MBH (see text; intermediate
solid line), isolated field stars (\citet{abt+02}; thin solid line)
and young cluster stars ($h$ and $\chi$ Persei, \citet{str+05};
dashed line).}
 
\end{figure}

We point out that the rotational velocity distribution of stars both
isolated and in binaries is very wide spread (see fig. \ref{f:cdfs}),
and therefore some statistics are required to test this signature
and the rotational velocity of a single star can not pinpoints to
its origin as a single or a binary star (many of the stars formed
in isolation are quite slow rotators \citep{abt+04}, where as some
of the binary evolved stars are very fast rotators). Using a Kolmogorov-Smirnov
test we find out that $\gtrsim25$ B MS HVSs are required, on average,
to be able to differentiate between these distributions with a $\ge95\%$confidence
level, if all these HVSs are taken from the same distribution (either
all evolved in binaries or all evolved in isolation). 

%

Binaries formed close to the MBH in the GC are soft binaries and would
shortly after be disrupted due to perturbing encounters with other
stars. Consequently the binary components, now single stars, should
also have lower rotational velocities, on average, similar to other
stars formed in binaries. Since the binary fractions of stars are
high (e.g. $>70\%$ for B stars in young clusters; \citealt{kob+06,kou+07}),
many, probably most, of the A and B MS stars in the GC are expected
to have formed in binaries as slow rotators, and later on become single
stars. If HVSs were ejected due to the SBH or IMBH kick scenarios,
most of them are therefore expected to be relatively slower rotators.
Still, a non-negligible fraction of the stars are formed as isolated
stars and possibly be faster rotators. 

We can construct the rotational velocity distribution of the currently
single stars populations close to the MBH ($<0.01$pc). The stellar
population in this region is composed of single stars formed in isolation
and of single stars originally formed in binaries that have evaporated
(with the appropriate fractions). The constructed rotational velocities
distribution is the combination of the single stars and binary stars
rotational velocities with the appropriate weights, that depend on
the binary fraction in the population. To be conservative we take
a lower limit on the initial binary fraction of $\sim35\%$ \citep{abt83}
where one should recall that each evaporated binary contributes two
stars to the combined single stars population. In other words, any
star chosen from our constructed rotational velocities distribution
in the GC has a $35\cdot2/(35\cdot2+65)=0.51$ probability to originally
form in a binary and therefore have a rotational velocity chosen from
the binary stars distribution, although it is currently a single star.
Again using a Kolmogorov-Smirnov test we find out that $\gtrsim100$
B MS HVSs are required, on average, to be able to differentiate between
these distributions with a $\ge95\%$confidence level, if all these
HVSs are taken from the same distribution (either all from the constructed
distribution for the stellar population in the close environment of
the MBH, or all evolved in binaries). Given the small number of HVSs
observed and inferred to exist, such a signature for the HVSs origin
is unfortunately quite weak (even weaker if a higher binary fraction
is assumed). 

Recently \citet{str+05} and \citet{wol+07} have shown that the rotational
velocity distribution in denser environments lack the cohort of slow
rotators, thus showing very different rotational velocity distribution
than field stars. Given these observations and our poor knowledge
on the star formation environments in the GC (both close to the MBH
and further out), it would be difficult to use the rotational velocities
of HVSs as a tracer for their ejection scenario. If the B MS stars
close to the MBH have formed in a dense environment (as would be expected
if they formed, e.g. in a stellar disk close to the MBH; \citealt{pau+06})
they are expected to be relatively fast rotators, i.e. opposite to
the expected distribution as discussed above. We conclude that the
rotational velocities of A and B MS HVSs are strongly dependent on
the formation environment of these stars, but are most likely not
good tracers for the ejection scenario of HVSs. Data on the rotational
velocity distribution of stars close to the MBH and further away,
may be an important clue for our understanding of their ejection mechanism,
but even in that case too large statistics may be required for them
to be used as a signature for the HVSs ejection scenario.

\subsection{On the asymmetric velocity distribution of observed hypervelocity
stars}

Current observation of HVSs detect B type stars of limited magnitude.
Such HVSs could either be MS B stars ($3-4\, M_{\odot}$; possibly
blue stragglers) or hot BHB stars. The velocity distribution of HVSs
shows a marked asymmetry between HVSs with much more HVSs with positive
Galactocentric radial velocities than HVSs with negative ones \citep{bro+07b}.
This was suggested to infer that the observed HVSs have short lifetimes,
and therefore bound HVSs are too short lived to be observed returning
with negative radial velocities \citep{bro+07b,kol+07,sve+07}. 

If HVSs are ejected continuously, such as in the ejection scenarios
of the binary disruption by a MBH or scattering by SBHs, then bound
HVSs ejected at earlier times could now be observed returning with
negative radial velocities. In this cases no asymmetry in the HVSs
velocity distribution should be observed (up to the escape velocity
from the galaxy, above which no returning stars are expected at any
time). Consequently the observations of asymmetry may raise a grave
problem for these scenarios, unless there is a special physical reason
for ejecting stars with short lifetimes. One explanation could be
related to the survival probability of binaries in the GC.

Hot BHB stars have been suggested to form through evolution in binaries 
and may have high binary fraction, similar to sdB stars 
\citep{pet+02a,pet+02b}. If this is the case, then the fast evaporation 
of binaries close to the MBH in the GC would exclude the formation 
of such stars (as well as sdB stars) and no such BHB HVS would be ejected
in the SBHs kick or IMBH inspiral scenarios (see discussion in section 
3.1).  In this case Hot BHB stars
would be very rare in the population of HVSs, and therefore
all or most of the observed HVSs are B MS stars, that naturally have
short lifetimes consistent with the asymmetric velocity distribution 
of observed HVSs. Alternatively, even if some of the HVSs are hot BHB
stars (e.g. from the binary disruption scenario, see section 3.1), 
they had to be ejected
only after they evolved to this stage, and therefore their propagation
time as HVSs is limited to their lifetime at this phase, which is
short (a few $10^{8}$ yrs) and comparable to that of MS B stars. In
both cases, an asymmetric velocity distribution of the HVSs would
be expected. 

HVSs could also be blue straggler stars (which would possibly give
them longer propagation times, and therefore different observable
velocity asymmetry). In this case the same arguments could be introduced
as for the hot BHB stars. Since the evolution of blue stragglers is
also through mass transfer in binaries (or stellar collisions, however,
this would not happen for an ejected star), and most, if not all of
the field blue stragglers are in binaries (\citealt{car+01}; note,
however, that these refer to lower mass blue stragglers), we may expect
to see only blue straggler HVSs that have been ejected already after
they evolved to this phase. Such stars are practically indistinguishable
from regular MS stars, and their lifetime at this stage is as short.

Another possibility for explaining HVSs velocity asymmetry is the
case of a limited time-span for the ejection of HVSs, such as expected
during an IMBH inspiral in the GC; i.e. a short lived discrete event,
and not long lived continuous process such as discussed above. In
this case stars are expected to be ejected only during the limited
and relatively short timescale at which the IMBH could eject HVSs
(unless several such inspiral events happened). Such timescale could
be as large as $10^{8}$yrs \citep{loc+07}, which could marginally
fit the observed ejection time span of the unbound HVSs (fig. 8 in
\citealt{bro+07b})%
\footnote{Notice, however, that scattering of stars by massive perturbers such
as giant molecular clouds and clumps could shorten the inspiral time
of the IMBH considerably \citep{per+08}.%
}. 

Recently the possibility of rare massive binary encounter in dense
young clusters \citep{gva+07} have been suggested for ejecting HVSs.
Such a process is also a continuous process which should have similarly
lead to a symmetric velocity distribution. However, in this scenario
mostly massive stars (and hence short lifetimes) are expected to be
ejected, which could explain the lack or high velocity returning stars.
However, it is not at all clear whether the necessary conditions in
such young clusters exist, and whether the frequency of such rare
strong encounters could explain the observed population of HVSs (and
especially unbound HVSs) to begin with (see \citealt{per08b} for
a short discussion on this).

We conclude that the currently observed B type HVSs are most likely
MS B stars, and suggest that hot BHB HVSs could only be produced in
the binary disruption scenario. However, even in the later scenario
these are not expected to be frequent.

\section{Summary\label{sec:Summary}}

In this study we have explored some dynamical and evolutionary constraints
on the nature and origin of HVSs and of the stellar population in
the GC. Hypervelocity stars are thought to be ejected through dynamical
interactions near the MBH in the GC. Three scenarios have been suggested
for their ejection; a disruption of a binary star by the MBH, scattering
by an intermediate mass BH which inspirals to the MBH or scattering
by stellar BHs in the close region of the MBH. In the binary disruption
scenario HVSs originate only from binaries, where most of them evolved
far from the MBH ($>2\, pc$). In the scattering scenarios by an intermediate
mass or stellar BHs most HVSs are single stars scattered from a close
region near the MBH ($<0.01$ pc from it). Given the differences between
them, the ejection scenarios of HVSs are expected to involve different
stellar populations in the GC. We have used dynamical and evolutionary
arguments together with current observations regarding the stellar
population in the GC to constrain the nature and origin of HVSs. We
have shown that the IMBH inspiral scenario requires too many main
sequence B stars to exist close to the MBH ($<0.01$ pc). Scattering
by SBHs are also not likely to be consistent with the observed population
of B stars in the Galactic center, although this scenario can still
be compatible with observations under extreme conditions. The binary
disruption scenario is still consistent with current observations. 

Due to the conditions close to the MBH most binary star systems are
not expected to survive for long in this region. Consequently unique
stellar populations that require a long evolution in a binary, such
as subdwarf (and possibly hot blue horizontal branch) B stars, blue
stragglers, Am stars and other populations are not expected to be
ejected as HVSs in the SBHs kicks or IMBH inspiral scenarios. In the
binary disruption scenarios the binaries involved originate much further
from the MBH where they could survive longer, and therefore HVSs of
these unique stellar population are not excluded, although their rates
might be quenched because of their shortened evolution in the binary
systems. Conversely, stellar populations that are not frequently observed
in close binaries such as required in the binary disruption scenario
(e.g. Be stars, A4-F2 type stars) are not expected to be ejected as
HVSs, or to be captured close to the MBH in this case, but they can
still possibly be ejected in the SBHs kicks scenarios. We also show
that these arguments suggest that signatures for HVSs origin such
as hypervelocity binaries and slow rotating HVSs may be much weaker
than expected and may require large statistics.

\acknowledgements
I would like to thank Warren Brown, Brad Hansen, Tal Alexander and
Uli Heber for helpful discussions and references. I would also like
to thank the Israeli Commercial \& Industrial Club for their support
through the Ilan Ramon scholarship.

\end{document}